# Unexpected large electrostatic gating by pyroelectric charge accumulation

Yicheng Mou[1#], Qi Liu[1#], Jiaqi Liu[1], Yingchao Xia[1], Zejing Guo[1], Wenqing Song[1], Jiaming Gu[1], Zixuan Xu[1], Wenbin Wang[1,2], Hangwen Guo[1,2], Wu Shi[1,2*], Jian Shen[1,2,3,4,5], Cheng Zhang[1,2*]

[1] State Key Laboratory of Surface Physics and Institute for Nanoelectronic Devices and Quantum Computing, Fudan University, Shanghai 200433, China
[2] Zhangjiang Fudan International Innovation Center, Fudan University, Shanghai 201210, China
[3] Shanghai Research Center for Quantum Sciences, Shanghai 201315, China
[4] Collaborative Innovation Center of Advanced Microstructures, Nanjing 210093, China
[5] Shanghai Branch, CAS Center for Excellence and Synergetic Innovation Center in Quantum Information and Quantum Physics, Shanghai 201315, China

[#] These authors contributed equally to this work.
[*] Correspondence and requests for materials should be addressed to W. S. (E-mail: shiwu@fudan.edu.cn) and C. Z. (E-mail: zhangcheng@fudan.edu.cn).

## Abstract

Pyroelectricity refers to the accumulation of charges due to changes in the spontaneous polarization of ferroelectric materials when subjected to temperature variations. Typically, these pyroelectric charges are considered unstable and dissipate quickly through interactions with the external environment. Consequently, the pyroelectric effect has been largely overlooked in ferroelectric field-effect transistors. In this work, we leverage the van der Waals interface of hBN to achieve a substantial and long-term electrostatic gating effect in graphene devices via the pyroelectric properties of a ferroelectric $LiNbO_3$ substrate. Upon cooling, the polarization change in $LiNbO_3$ induces high doping concentrations up to $10^{13}$ cm$^{-2}$ in the adjacent graphene. Through a combination of transport measurements and non-contact techniques, we demonstrate that the pyroelectric charge accumulation, as well as its enhancement in electric fields, are responsible for this unexpectedly high doping level. Our findings introduce a novel mechanism for voltage-free electrostatic gating control with long retention.

Key words: electrostatic gating, pyroelectric effect, ferroelectricity, graphene, surface acoustic waves

## Introduction

The unique properties of non-centrosymmetric crystals, particularly their piezoelectric and pyroelectric responses, have increasingly attracted attention for their ability to convert mechanical and thermal energies into electric signals[1–4]. This capability underpins a wide array of applications, from surface acoustic wave (SAW) filters and oscillators, to thermal energy harvesters and imaging devices. Among these materials, ferroelectric crystals stand out for their spontaneous polarization below the Curie temperature and the reversible polarization under an external electric field, making them invaluable for non-volatile memory and memristive devices[5–8]. A prominent example is the ferroelectric field-effect transistor (FeFET). Due to its intrinsic nonvolatile characteristics, low energy consumption, and compatibility with modern silicon-based CMOS technology[9], FeFET has emerged as a leading candidate for next-generation memory devices[10–13]. Furthermore, FeFET holds promise for in-memory computing and artificial neural networks, where both memory and computation are integrated within a single device, overcoming the limitation of von Neumann architecture[14].

In a FeFET, the polarization state of the ferroelectric layer modulates the carrier density in the underlying semiconductor channel, thereby controlling its conductance. This modulation arises from the strong electric field generated by the ferroelectric polarization, which can be switched between two stable states, enabling binary data storage[15–17]. While most research has focused on the electric-field-driven switching of ferroelectric polarization, the role of temperature changes, mediated by pyroelectricity, has been largely overlooked in FeFET. When the spontaneous polarization strength of the ferroelectric layer shifts due to temperature variation or thermal fluctuations, pyroelectric charges accumulate on the material surface, generating a voltage across the crystal[18]. However, these excess surface charges are typically unstable in ambient environments. They tend to attract and trap free charges with opposite sign from surroundings, leading to discharge processes, such as ion movement in the air, electron migration within the crystal, or leakage currents through external circuits. Consequently, this second-order effect has been considered negligible and largely ignored in FeFET research.

In this work, we demonstrate a long-term stabilization of pyroelectric charges of ferroelectric LiNbO$_3$ (LNO) substrates, unlocking their potential for sustained large electrostatic gating. As the temperature decreases, polarization changes in LNO result in high doping concentrations of up to $10^{13}$ cm$^{-2}$ in the adjacent hBN-encapsulated graphene without degrading the carrier mobility. To uncover the physical origin, we employed a combination of transport measurements and SAW attenuation techniques. Our results reveal that the unexpectedly high doping levels arise from charge accumulation due to the pyroelectric effect, which is further amplified by electric field. The high-quality van der Waals interface of hBN encapsulation allows for long-term storage of the pyroelectric charges, forming a nonvolatile electrostatic gating with extended retention, all without requiring external voltage. This discovery introduces a novel approach for controlling electrostatic gating, which is crucial for low-temperature quantum material studies and the development of temperature-sensitive electronic switches.

## Device structure and carrier density modulation

As depicted in the ferroelectric phase crystal structure in Fig. 1a, the ferroelectricity of LNO arises directly from ion displacement[19]. Below its ferroelectric Curie temperature (about 1210 °C), LNO consists of planar sheets of oxygen atoms in a distorted hexagonal close-packed arrangement.

Lithium, niobium, and vacancies equally occupy the octahedral interstices within this structure. Along the $c$-axis, elastic forces shift the cations (Li and Nb ions) slightly away from the center of anions (oxygen octahedra) at temperatures below 1210 °C, resulting in spontaneous polarization along the $c$-axis[20,21]. The polarization field generates bound charges that accumulate on the top and bottom surfaces of the LNO substrate. In industrial and research applications, special cutting orientations for LNO wafer, such as Z-cut, Y-cut, 64°Y-cut, and 128°Y-cut, are commonly employed, as illustrated in Fig. 1b. For the Z-cut wafer, the surface normal aligns with the $+z$-axis ($c$-axis), while in the Y-cut wafer, it points along the $+y$-axis ($b$-axis). For the 128°Y-cut wafer, the surface normal forms an angle of approximately 128° with the $y$-axis.

To investigate the electrostatic doping induced by pyroelectricity, we fabricated graphene field-effect transistors on top of LNO substrates with different orientations. The LNO substrates were commercially purchased and ferroelectrically polarized initially. The device schematic used in this study is shown in Fig. 1c. The graphene sheet was encapsulated between the top and bottom hBN layers by a dry transfer method[22–24]. The heterostructure was then transferred onto the LNO substrate and patterned into a Hall-bar geometry (see details in Supporting Information). Apart from electrodes contacted with graphene, there are additional electrodes on top of the top BN layer and beneath the LNO substrate, serving as the top and bottom gates, respectively.

Magnetotransport measurements were carried out in variable temperature inserts with a superconducting magnet in a dry cryostat. Initially, we measured the four-terminal longitudinal resistance $R_{xx}$ and Hall resistance $R_{yx}$ of hBN-encapsulated graphene on both 128°Y-cut and Y-cut LNO substrates. Here the out-of-plane component of ferroelectric polarization is non-zero for the 128°Y-cut LNO substrate, while absent for the Y-cut one. The results shown in Fig. 1d-f reveal striking differences between samples on the two substrates. Figure 1d shows the normalized resistance as a function of temperature. When cooling down to 2 K, $R_{xx}$ on the 128°Y-cut LNO substrate (denoted as $R_{xx}$-128°Y) decreases much more rapidly than $R_{xx}$-Y. At 2 K, $R_{xx}$-128°Y is only 33.3 Ω, over one order magnitude smaller than that from $R_{xx}$-Y (> 600 Ω). By tracking the Hall resistance (Fig. 1e), we find that the Hall coefficient $R_H$-128°Y is only -60 Ω/T at 2 K (corresponding to a heavily-doped electron concentration of $1.05 \times 10^{13} \mathrm{cm}^{-2}$), while $R_H$-Y is -1025 Ω/T (corresponding to a concentration of $6.09 \times 10^{11} \mathrm{cm}^{-2}$). The latter is close to that of conventional graphene on Si/SiO$_2$ substrates [25,26]. The calculated Hall mobility of graphene on the 128°Y-cut LNO substrate is $3.61 \times 10^4 \mathrm{cm}^2/(\mathrm{V \cdot s})$ at 2 K, indicating the maintenance of high sample quality even at elevated carrier densities. In contrast, $R_H$-128°Y and $R_H$-Y have similar value at 300 K as shown in Fig. S2a. We further measured Shubnikov-de Haas (SdH) oscillations under perpendicular magnetic fields to directly probe the Fermi surface. As shown in Fig. 1f, the dense oscillation pattern for the 128°Y sample indicates a relatively large Fermi surface with a frequency of 77 T. We can extract carrier concentration based on the oscillation frequency as $n_{\mathrm{SdH}} = 0.74 \times 10^{13} \mathrm{cm}^{-2}$, close to the above Hall carrier density.

## Electric transport measurements

To elucidate the electrostatic doping for the 128°Y sample, we conducted systematic temperature- and field-dependent transport measurements. By measuring the Hall resistivity at $\pm 1$ T, we extract the temperature dependence of carrier density $n_{\mathrm{Hall}}$. The change in carrier density $|\Delta n| = |n_H(T) - n_H(300 \mathrm{~K})|$ as a function of temperature is presented in Fig. 2a. In contrast to the weak temperature dependence of the Y-cut case, the $|\Delta n| - T$ curve for the sample on the 128°Y-cut substrate shows a continuous and significant increase, indicating pronounced electron doping as the temperature decreases. We also fabricated another graphene device with top gate on the 128°Y-cut substrate. Its transfer curves shown in the inset of Fig. 2a show a systematic shift of charge-neutral point (CNP) with temperature (from $V_{GS} = -3$ V at 195 K to $V_{GS} = -7$ V at 182 K, and disappear below 160 K), in consistent with the cooling-induced electron doping behavior presented above. On the contrary, the CNP of graphene on Y-cut substrate remains unchanged even at very low temperatures down to 2 K (refer to Fig. S2b). Figure 2b compares the Hall resistance of two graphene devices fabricated on opposite surfaces of the 128°Y-cut substrate at 2 K, in which the ferroelectric polarization directions are opposite. It results in a reversed hole doping effect for the -$z$-axis-polarized substrate. The $n_{Hall}(T)$ curve of this sample also shows an opposite trend to that on the +$z$-axis-polarized substrate in Fig. 2a, with an $n$-to-$p$ carrier type transition occurring around 290 K (refer to Fig. S2c).

These compelling results indicate that the doping stems from the pyroelectric properties of the LNO substrate. As illustrated in Fig. 2c, spontaneous polarization in LNO at room temperature results in bound charges accumulating on both surfaces. These bound charges are easily screened by free charges during fabrication and measurement in air, which explains the absence of doping at 300 K. However, as the temperature decreases, the pyroelectric effect generates excess bound charge due to increased ferroelectric polarization. In a vacuum chamber, without external free charges to neutralize them, the polarization electric field penetrates the graphene channel, attracting charges with the opposite sign and causing significant doping.

This mechanism also explains the lack of doping in Y-cut LNO, which has no out-of-plane polarization component, and the reversal of doping polarity with opposite polarizations. The inset of Fig. 2d shows the carrier density variation $\Delta n$ around 300 K, further supporting the pyroelectric origin of the electrostatic gating, with opposite charge doping observed during heating and cooling. Moreover, this voltage-free electrostatic gating is non-volatile, as the doped charges dissipate very slowly. As shown in Fig. 2d, the plot of $R_{xx}$ of a device on 128°Y LNO versus time reveals a gradual change in resistance. By fitting this variation with an exponential decay function that accounts for the reduction in excess bound charges, we extract a long retention time as $\tau \approx 1.13 \times 10^7$ s.

**Non-contact measurements**

To directly test the proposed pyroelectric-induced electrostatic gating mechanism, we employed a non-contact SAW attenuation technique, which utilizes the amplitude attenuation ($\Gamma$) and velocity shift ($\Delta v/v_0$) of sound wave to probe the electronic state in a contactless manner[27]. Taking advantage of the piezoelectric nature of the LNO substrate, strain can be induced in the presence of an external electric field. By sending a radio-frequency electric excitation to interdigital transducers fabricated on LNO, SAW can be generated, accompanied by strain and piezoelectric fields[28]. When SAW interacts with conductive materials on the piezoelectric surface, carriers partially screen the piezoelectric field through a relaxation process, leading to dissipations and shifts in SAW velocity, which are directly related to the material conductivity. This technique has been utilized in the study of acoustoelectric effects in AlGaAs/GaAs two-dimensional electron gas[27,29–34] and more recently, in two-dimensional materials[24,35–38].

Here a SAW resonator was used to enhance the interaction between acoustic wave and the graphene channel as shown in Fig. 3a. The excited SAW was reflected back and forth by a series pairs of short-circuited electrodes, forming a Fabry-Perot cavity with a high quality factor. The fabrication and characterization of the SAW resonator are described in detail in Supporting Information. During the measurement, the RF signal generated by an ultra-high frequency lock-in amplifier (Zurich Instrument UHFLI) was input into one of the IDTs of the resonator (IDT1), and the output from the opposite IDT (IDT2) was collected by the same lock-in with a low-noise amplifier (LNA) and a bandpass filter. Typically, IDT1 was driven at 1 mV$_{rms}$ with a frequency of 222 MHz. The measured phase difference $\Delta\theta$ (defined as $\theta(B) - \theta(0\,T)$) and the RF amplitude $|V_{RF}|$ relate to the SAW velocity shift and attenuation through the following equations:

$$\frac{\Delta\theta}{360°} = \frac{Q}{\pi}\frac{A_s}{A_c}\frac{\Delta v}{v_0}\,, \quad (1)$$

$$\frac{|V_{RF}|}{|V_{RF0}|} \propto e^{-\Gamma L}\,, \quad (2)$$

where $Q$ is the quality factor of the Fabry-Perot resonator, $|V_{RF0}|$ is the amplitude at zero magnetic field, $L$ is the sample length, and $A_s$ and $A_c$ are the areas of the sample and cavity, respectively.

We performed SAW attenuation and velocity shift measurements for BN-encapsulated graphene devices with and without contacting electrodes in a variable temperature insert equipped with a custom-designed RF probe. As illustrated in Fig. 3b, for the pyroelectric-induced electrostatic gating mechanism, doping only occurs when the graphene sample is connected to the ground by electrodes during the cooling process. In contrast, other doping mechanisms, such as interface charge transfer, remain unaffected by the presence of contacting electrodes. Figure 3c-d shows the $\Delta\theta$ and $|V_{RF}|$ / $|V_{RF0}|$ results measured on a device without and with contacting electrodes. Prominent quantum oscillations are observed in both devices but with distinct oscillation frequencies. The uncontacted device shows a small Fermi surface with an oscillation frequency around 15.4 T, while the contacted one has a frequency around 83.3 T. In Fig. S2d, we present the transport data of a graphene device on the Y-cut substrate, in which the interface pyroelectric effect is absent due to the in-plane polarization. The Y-cut device shows clear quantum Hall effect with an oscillation frequency around 6.4 T, close to that of the uncontacted one in Fig. 3c. Following similar analysis of SAW attenuation in previous studies[27,39], we attribute the peaks of $\Delta\theta$ and the valleys of $|V_{RF}|$ / $|V_{RF0}|$ in Fig. 3c to the quantum Hall plateau of $|\nu| = 6$ and $|\nu| = 10$, respectively. In Fig. 3e, the Landau fan diagram extracted from the peaks of $\Delta\theta$ further confirms a larger Fermi surface and higher Landau indices in the contacted device. The quantum oscillations in resistance in Fig. 3f (with the device image inset in Fig. 3d) agree with the SAW attenuation results, and the inset of Fig. 3f shows the corresponding conductivity $\sigma_{xx}$ versus $1/B$. Figure 3g compares carrier densities derived from Hall effect and quantum oscillations across various samples and contacting conditions. High doping occurs only in the contacted device on LNO substrates with out-of-plane polarization. These findings suggest that electrostatic gating, rather than direct interface charge transfer, is responsible for the observed high doping in hBN-encapsulated graphene on LNO substrates.

## Electric field enhanced polarizations

For a quantitative analysis of the doping concentration, we extract the change in spontaneous polarization $\Delta P_s$ by integrating the pyroelectric coefficient $p = dP_s/dT$ (See Supporting Information for details)[18]. As temperature decreases, $\Delta P_s$ shows a continuous increase, eventually saturating, which aligns closely with the trend observed in $|\Delta n|$. For comparison, we also measured the pyroelectric response of Z-cut LNO. As expected, the Z-cut substrate exhibits a larger pyroelectric current due to its polarization field being oriented normal to the surface (refer to Fig.

S3c). As shown in Fig. 4a, the obtained $\Delta P_s$ of Z-cut LNO is slightly larger than that of the 128°Y-cut substrate. Correspondingly, the $|\Delta n|$ on Z-cut substrates can reach higher values than that on 128°Y-cut as shown in Fig. S2e, which matches with the measured $\Delta P_s$.

By quantitatively comparing $\Delta P_s$ and $\Delta n$, we find that the observed doping concentration substantially exceeds that predicted by the polarization change. The polarization change $\Delta P_s = 0.92 \ \mu C/cm^2$ for the 128°Y-cut substrate at 2 K, corresponding to a doping level of $0.57 \times 10^{13}$ $cm^{-2}$, only accounts for roughly half of the observed $\Delta n$ in Fig. 2a. The extra doping part of $\Delta n$ can be removed by modifying the substrate from contacted to floated configurations (data in Fig. 1, 2, and 3 were acquired in contacted configuration). The doping level $\Delta n$ for the floated substrate is around $0.4 \times 10^{13}$ $cm^{-2}$, which is smaller than the above value of $\Delta P_s$. The comparison of temperature dependence of $\Delta n$ for the substrate-contacted and floated configurations is summarized in Fig. 4b, showing divergence at low temperatures. The inset of the Fig. 4b presents $\Delta n$ versus $\Delta P_s$ for both configurations, with the theoretical values depicted by the black dashed line, following the equation $\Delta n = \Delta P_s/e$. Here $e$ is the elementary charge. Figure 4c shows the Hall resistance and the oscillation frequency spectrum (inset) at 2 K for both configurations. The slope of the substrate-floated sample is about -140 $\Omega$/T, corresponding to a carrier density of $4.5 \times 10^{12}$ $cm^{-2}$, contrasting the -60 $\Omega$/T Hall coefficient observed for the substrate-contacted condition. The inset shows the fast Fourier transform of SdH oscillations, revealing frequencies of 35 T and 77 T, supporting the enhanced doping effect for the substrate-grounded configuration. The mechanism behind the enhanced doping when the substrate is grounded is illustrated schematically in Fig. 4d. The $\Delta P_s$ value obtained from pyroelectric current measurement corresponds to the change in intrinsic ferroelectric polarization without external fields. However, when the LNO substrate is grounded, positive charges are generated on the grounding electrode. It contributes to an additional electric field that strengthens the ferroelectric polarization in LNO, therefore resulting in an enhanced $\Delta P_s$.

## Discussions and summary

Our discovery of the pyroelectric-induced electrostatic gating effect represents a novel and important advancement in the field of carrier modulation in quantum materials. Here the electrostatic doping is achieved through the capacitive coupling between the pyroelectric charges and the channel material, which differs from conventional charge transfer effect induced by ferroelectric polarization[40,41]. The latter has been extensively used in various device applications such as FeFET[15,16]. This voltage-free gate mechanism, capable of achieving a high carrier concentration ($\sim 10^{13} cm^{-2}$), introduces a new approach to controlling electronic properties. Importantly, this doping level can be further amplified with the application of an external electric field, enabling access to even higher carrier densities. In the context of quantum materials research, precise control over carrier density is critical for uncovering new physical phenomena, leading to the discovery of novel phenomena such as superconductivity[42,43], room-temperature ferromagnetism[44], and phase transitions[45,46]. Our technique, utilizing the intrinsic pyroelectric properties of ferroelectric substrates, opens new pathways for such discoveries by providing a clean, non-invasive method of charge doping, without the need for external voltages or chemical modifications.

In addition to fundamental research, the pyroelectric-induced gating effect offers promising avenues for novel device applications. By closely linking temperature variations to electrostatic gating, it provides opportunities for designing thermal memory devices or temperature-controlled switches that can operate stably under ambient conditions. Furthermore, the ability to locally control gating by selectively depositing screening layers on the LNO substrate enables the creation of *p-n* junctions and spatially distinct *p*- and *n*-doped regions on a single platform. This local control over carrier density could lead to innovative device architectures, providing flexibility in designing advanced electronics and optoelectronics.

In summary, we have introduced a new technique for carrier doping in low-dimensional materials, exploiting the large change in polarization ($\Delta P_s$) of ferroelectric substrates as temperature varies. This approach enables the modulation of carrier densities exceeding $10^{13}$ cm$^{-2}$, while preserving the material's intrinsic quality by avoiding extrinsic impurities or structural damage. Our work not only contributes a novel electrostatic gating mechanism but also provides a versatile platform for patterning superlattice potentials and customizing the electronic properties of low-dimensional systems.

## Supporting Information
Device Fabrication & Sample Characterization; Other Transport Measurement Results; Pyroelectricity Measurements; Fabrication and Characterization of Fabry-Perot SAW Resonators.

## Acknowledgements

We would like to thank Lin Fu, Jinfeng Zhai and Bofan Xu for helpful discussions and experimental support. This work was supported by the National Key R&D Program of China (Grant No. 2022YFA1405700), the National Natural Science Foundation of China (Grant No. 92365104 and 12174069), and Shuguang Program from the Shanghai Education Development Foundation. Part of the sample fabrication was performed at Fudan Nano-fabrication Laboratory.

## Author contributions

Y.M. and Q.L. contributed equally to this work. C.Z. supervised the overall research. Y.M., Q.L., J.L., and Y.X. fabricated the devices assisted by Z.G., W.Q.S., J.G., and Z.X. Y.M. and Q.L. carried out the electrical transport and surface acoustic wave measurements. Y.M. performed pyroelectric coefficient measurements. Y.M. and J.L. performed Raman spectrum measurements. Y.M., Q.L., W.W., H.G., W.S., J.S., and C.Z. analyzed the data. Y.M., Q.L., and C.Z. prepared the manuscript with contributions from all other co-authors.

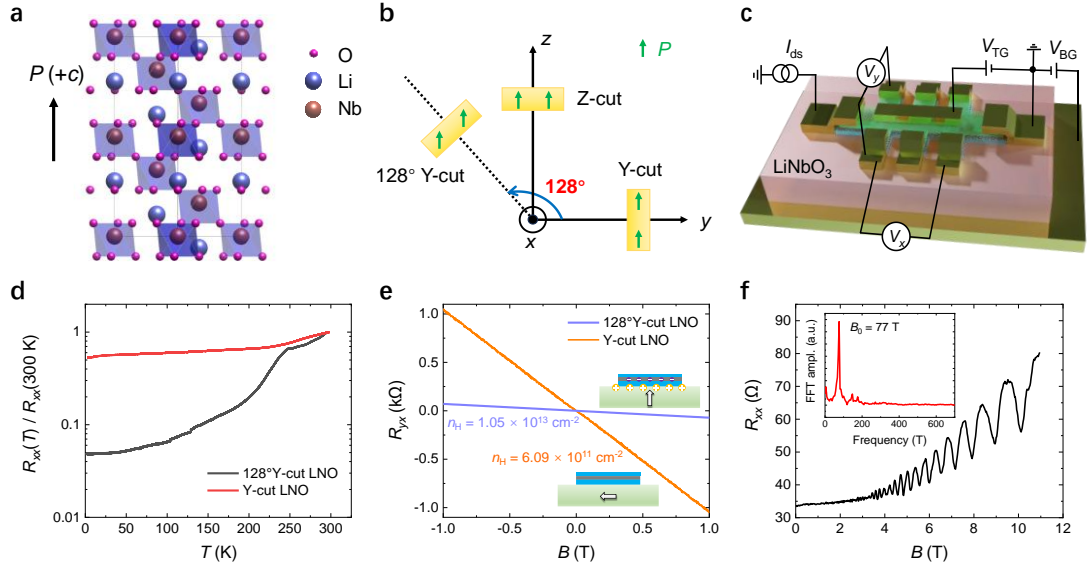

**Figure 1 | Device structure and carrier density modulation.** (**a**) The crystal structure of ferroelectric phase LNO with spontaneous polarization along the *c*-axis. (**b**) Schematic of different cut orientations of commercial single-crystal LNO substrates. The yellow rectangle denotes the side view of the wafer, the green arrow represents the spontaneous polarization field, and the *x*-axis points in the out-of-plane direction. (**c**) Device structure and measurement setup of the hBN-encapsulated graphene device, where $I_{ds}$ is the drain-source current, and $V_{TG}$ and $V_{BG}$ are the top and back gate voltages, respectively. (**d**) Comparison of normalized temperature-dependent resistances ($R_{xx}(T) / R_{xx}(300\ \mathrm{K})$) of two graphene devices on 128°Y-cut and Y-cut LNO$_3$ substrates, with one showing an order of magnitude decrease in resistance while the other only slightly decreases. (**e**) Comparison of Hall resistance of the graphene devices on 128°Y-cut and Y-cut LNO$_3$ substrates at 2 K, with the 128°Y-cut sample exhibiting high doping over $10^{13}$ cm$^{-2}$, while the Y-cut one remains low carrier density. (**f**) Magnetoresistance of the 128°Y-cut sample at 2 K. The inset shows the fast Fourier transform of SdH oscillations, revealing an oscillation frequency of $B_0 = 77$ T.

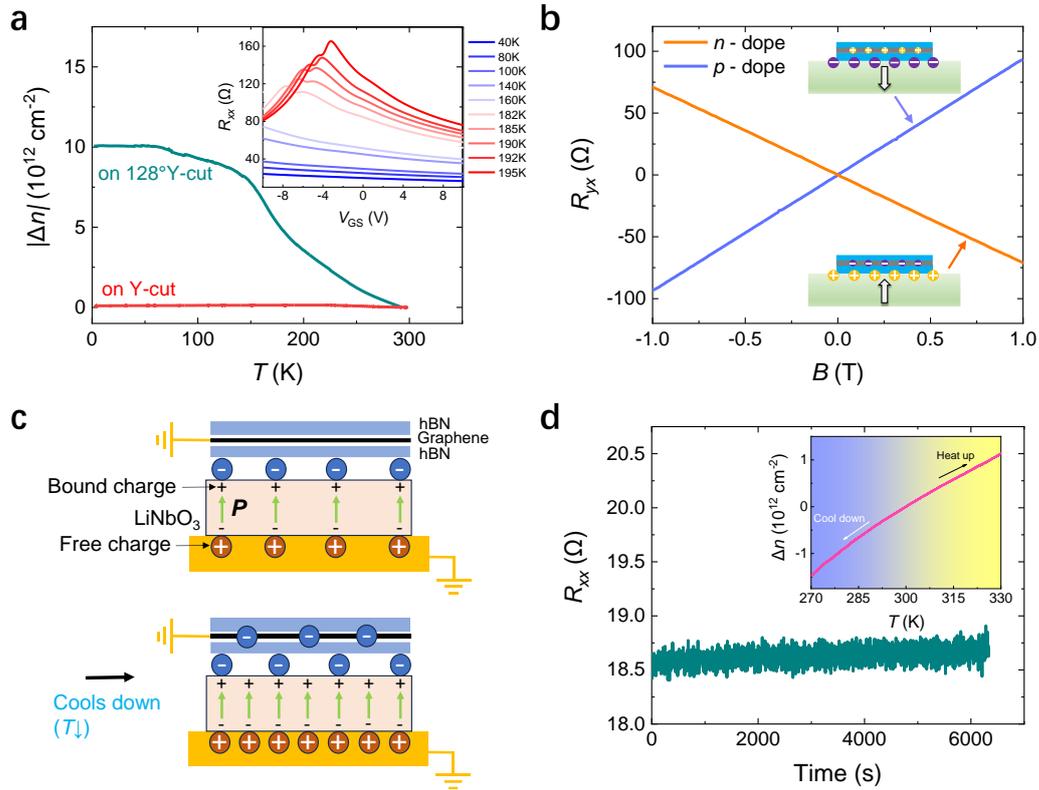

**Figure 2 | Electric transport results and mechanism illustration.** (**a**) Comparison of the absolute variation in carrier density between samples on 128°Y-cut and Y-cut LNO as a function of temperature. The inset presents transfer curves measured at temperatures from 40 K to 195 K using the hBN top gate, demonstrating a systematic shift of the CNP position with temperature. (**b**) Comparison of Hall resistance for graphene devices fabricated on opposite surfaces of 128°Y-cut substrates, showing opposite carrier types with similar concentrations. (**c**) Schematic illustration of the pyroelectric-induced electrostatic gating mechanism. Up: Device at room temperature, where the green arrow denotes spontaneous polarization, and the blue and orange circles represent external free charges from environment. Bottom: Device at a lower temperature, where the increased polarization due to pyroelectricity results in enhanced doping. (**d**) The time-dependent plot of $R_{xx}$ of a device on 128°Y-cut LNO showing the retention of electrostatic gating. The resistance of the 128° Y-cut sample slowly varies as time goes on, indicating a very slow release of charges. The inset shows the carrier density variation $\Delta n$ around 300 K, where opposite charge doping is observed during heating and cooling, confirming that the effect is induced by pyroelectricity.

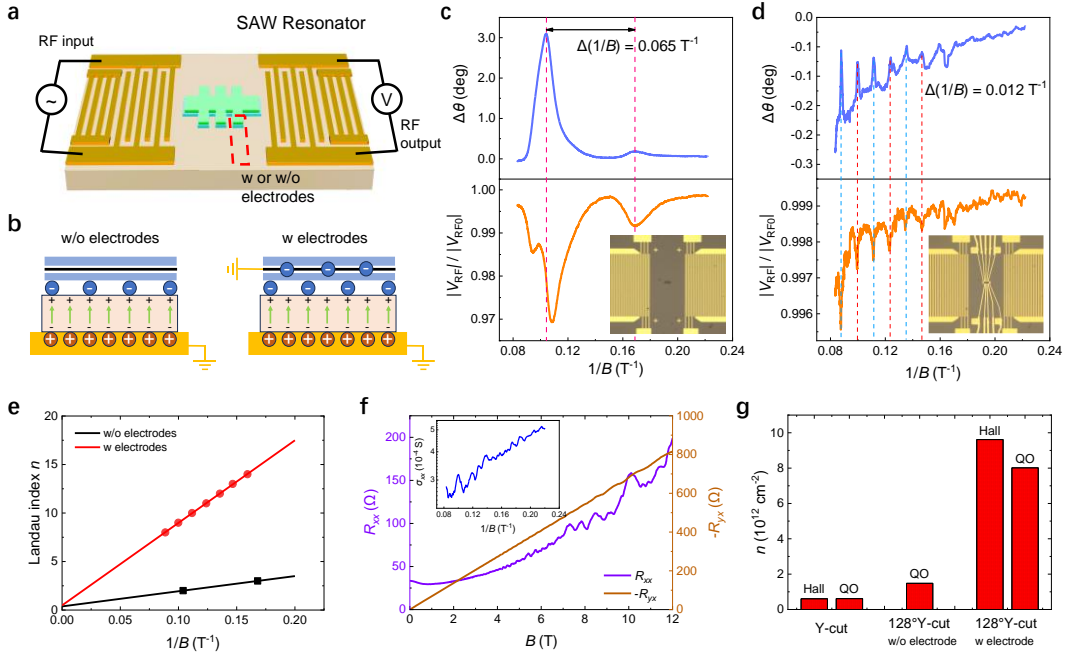

**Figure 3 | Results of surface acoustic wave attenuation and velocity shift measurements. (a)** Schematic of the Fabry-Perot SAW resonator device and measurement setup for SAW attenuation experiment. The resonator consists of a pair of opposite interdigital transducers and a series of shorted metal reflectors. **(b)** Illustration of different gating conditions for samples without and with electrodes. The metallic electrodes act as a doping source, injecting carriers into the graphene sheet. **(c)-(d)** SAW measurement results of samples without and with electrodes, respectively, as functions of $1/B$ at 2 K. Here, the phase difference $\Delta\theta$ is defined as $\Delta\theta(B) = \theta(B) - \theta(0\text{ T})$, which is related to the sound velocity shift. $|V_{RF}|/|V_{RF0}|$ corresponds to the SAW amplitude attenuation, where $|V_{RF0}|$ is the measured lock-in voltage at 0 T. Quantum oscillations are observed under both conditions, but with distinct frequencies. The insets show optical images of the same sample before and after contacted with electrodes. **(e)** Landau fan diagram derived from the phase shift $\Delta\theta$ in samples without and with electrodes. **(f)** Transport measurements of the sample in the SAW resonator as shown in **d**. The oscillation behavior and frequency match those observed in $\Delta\theta$ and $|V_{RF}|/|V_{RF0}|$. The inset presents the conductivity data as a function of $1/B$. **(g)** Comparison of carrier densities in graphene samples under different conditions, as determined from Hall effect measurements and quantum oscillations (QO), respectively.

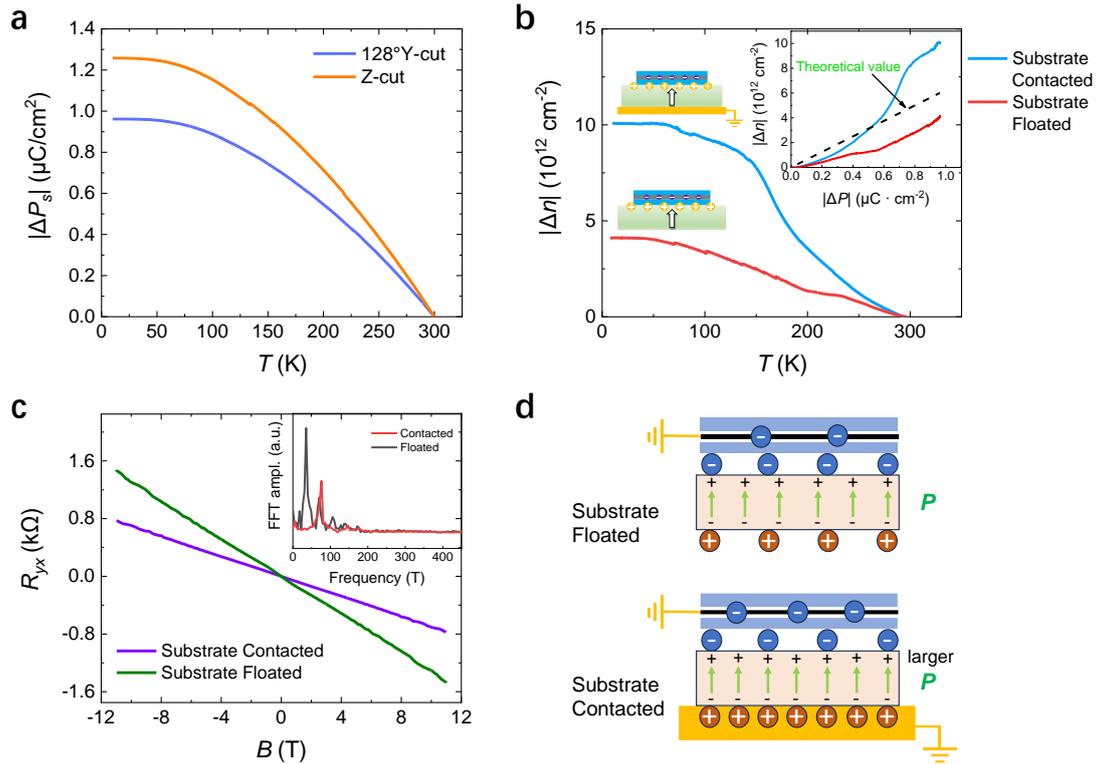

**Figure 4 | Evidence of the electric-field-enhanced polarization. (a)** Absolute values of spontaneous polarization variation of 128°Y-cut and Z-cut LNO versus temperature, calculated from the measured pyroelectric coefficients. Here, $\Delta P$ is defined by $\Delta P(T) = P(T) - P(300$ K$)$. **(b)** Comparison of $|\Delta n|$ versus temperature for the 128°Y-cut samples with the substrate grounded or floated. The inset shows the relation between $|\Delta n|$ and $|\Delta P|$, with the black dashed line indicating theoretical values. **(c)** The Hall resistance of the 128°Y-cut samples with substrate grounded or floated. The inset shows the fast Fourier transform of SdH oscillations of both conditions. **(d)** Illustration of polarization enhancement in the metal substrate contacted states.

TOC Graphic

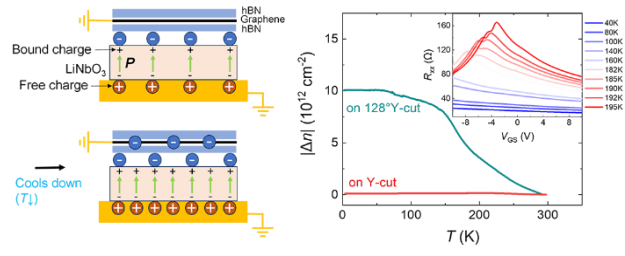